\documentclass[12pt]{article}
\usepackage{graphicx}
\usepackage{amssymb}
\usepackage{bm}% bold math
\usepackage[authoryear]{natbib}
\renewcommand{\appendix}{\par
        \setcounter{section}{0}
        \def\thesection{Appendix.}
}

\title{Impact of deviation from precise balance of spike-timing-dependent plasticity}

\author{Narihisa Matsumoto$^{1,2,3}$ and Masato Okada$^{2,3}$\\
$^1$Graduate School of Science and Engineering, Saitama University, \\
Saitama 338-8570, Japan\\
$^2$Lab. for Mathematical Neuroscience, RIKEN Brain Science Institute, \\
Saitama 351-0198, Japan\\
$^3$PRESTO, Japan Science and Technology Agency, Saitama 351-0198, Japan\\
Corresponding author \\
Narihisa Matsumoto, \\
Lab. for Mathematical Neuroscience, RIKEN Brain Science Institute, \\
2-1 Hirosawa, Wako-shi, Saitama 351-0198, Japan \\
Phone: +81-48-467-9664 \\
Fax: +81-48-467-9693 \\
E-mail: xmatumo@brain.riken.go.jp
}

\date{}

\begin{document}
\maketitle
\thispagestyle{myheadings}

\newpage
Impact of deviation from precise balance of spike-timing-dependent plasticity
\section*{Abstract}
Recent biological experimental findings have shown that synaptic
plasticity depends on the relative timing of pre- and post-synaptic
spikes and this is called spike-timing-dependent plasticity (STDP).
Many authors have claimed that a precise balance between long-term
potentiation (LTP) and long-term depression (LTD) of STDP is
crucial in the storage of spatio-temporal patterns.
Some authors have \textit{numerically} investigated the impact of an
imbalance between LTP and LTD on the network properties.
However, the mathematical mechanism remains unknown.
We \textit{analytically} show that an associative memory network
has the robust retrieval properties of spatio-temporal patterns, and
these properties make the network less vulnerable to any deviation from 
a precise balance between LTP and LTD when the network contains a 
\textit{finite} number of neurons.

\section*{Keywords}
spike-timing-dependent plasticity; long-term depression; long-term potentiation; retrieval properties; spatio-temporal patterns; statistical neurodynamics

\newpage
\section{Introduction}
Recent biological experimental findings have indicated that 
synaptic plasticity depends on the relative timing of pre- and 
post-synaptic spikes.
This relative timing determines whether long-term potentiation
(LTP) or long-term depression (LTD) is induced \cite[]{Bi98,
Markram97, Zhang98}.
LTP occurs when presynaptic firing precedes postsynaptic
firing by no more than about $20$ ms.
In contrast, LTD occurs when presynaptic firing follows postsynaptic firing.
A rapid transition between LTP and LTD takes place within a few milliseconds.
A learning rule of this type is called spike-timing-dependent
plasticity (STDP) \cite[]{Song00} or temporally asymmetric Hebbian
learning (TAH) \cite[]{Abbott99, Rubin01}.
The functional role of STDP has been investigated by many authors
\cite[]{Gerstner96,Kempter99,Abbott99,Munro00,Rao00,Song00,Rubin01,Levy01,Rossum01,Song01,Yoshioka02,Fu02,Karmarkar02,Matsumoto02}.
Some of these authors have claimed that the precise balance between
the LTP and LTD of STDP is crucial in the storage of spatio-temporal pattterns
\cite[]{Munro00,Yoshioka02,Matsumoto02}.
In our previous work, we analytically showed that STDP has the
same qualitative effect as the covariance rule when the balance is 
precisely maintained \cite[]{Matsumoto02}.
In the brain, though, it is inconceivable that such a balance will be
precisely maintained.
Some authors \textit{numerically} have investigated the impact of an
imbalance between LTP and LTD on network properties
\cite[]{Song00,Munro00,Yoshioka02}.
Munro and Hernandez showed that spatio-temporal patterns cannot
be retrieved in a noisy environment when there is no LTD
\cite[]{Munro00}. 
However, the \textit{mathematical} mechanism that accounts for this remains unknown.

In this paper, we \textit{analytically} investigate the retrieval
properties of spatio-temporal patterns in an associative memory
network where there is an imbalance between the LTP and LTD of STDP by using
a method of the statistical neurodynamics
\cite[]{Amari88,Okada96,Matsumoto02}.
Using this theory, we discuss the macroscopic behavior of the
network at a thermodynamic limit: the number of neurons is
infinite.
When the balance is precisely maintained, sequential patterns can
be stored in the network \cite[]{Matsumoto02}.
Also, when the mean deviation from the precise balance is $0$ and the
fluctuation is finite, patterns can be stored at a thermodynamic limit \cite[]{Matsumoto03}.
This implies that when the balance is maintained on average, patterns can be stored.
We also showed that when the mean is not $0$, patterns cannot
be stored since a cross-talk noise diverges at a thermodynamic limit.
However, previous work using computer simulation has shown that
the stored limit cycle is stably retrieved when the number of
neurons is \textit{finite} \cite[]{Munro00,Yoshioka02}.
In the brain, the number of neurons is considered to be
\textit{finite}.
Therefore, it is important to discuss this situation,
although no analytical works regarding this has been reported.
The purpose of the work reported in this paper was to \textit{analytically} investigate
the retrieval properties when the number of neurons is finite.
We found that a network containing a finite number of neurons
becomes robust regarding this deviation from a precise balance
between the LTP and LTD of STDP.
Since the number of neurons in the brain is considered to be
finite, our results might apply to the brain.

We introduce our model for the storage of spatio-temporal patterns
in section 2.
In section 3, we analytically discuss our model as it applies when the
balance between the LTP and LTD of STDP is precisely maintained and
when it is not precisely maintained, and in section 4 we investigate the properties of our model.
In section 5, we summarize this paper.
The Appendix presents a detailed derivation of the model's dynamical equations
of the model when the balance is precisely maintained.

\section{Model}
The model contains $N$ binary neurons with reciprocal
connections.
Each neuron takes a binary value: $\{0, 1\}$.
We define discrete time steps and the following rule for
synchronous updating:
\begin{eqnarray}
u_i(t) &=& \sum_{j=1}^N J_{ij} x_j(t), \label{eq.dynamics1} \\
x_i(t+1) &=& F( u_i(t) - \theta ), \label{eq.dynamics2} \\
F(u) &=& \left\{ \begin{array}{ll}
		1. & u \geq 0 \\
		0. & u < 0, \\
		\end{array} \right.
\label{eq.dynamics3} 
\end{eqnarray}
where $x_i(t)$ is the state of the $i$-th neuron at time $t$, 
$u_i(t)$ is the internal potential of that neuron, and $\theta$ is 
a uniform threshold.
If the $i$-th neuron fires at $t$, its state is $x_i(t) = 1$;
otherwise, $x_i(t) = 0$.
$J_{ij}$ is the synaptic weight from the $j$-th neuron to the
$i$-th neuron.
Each element $\xi^{\mu}_i$ of the $\mu$-th memory pattern
$\bm{\xi}^{\mu} = (\xi^{\mu}_1, \xi^{\mu}_2, \cdots, \xi^{\mu}_N)$
is generated independently by
\begin{equation}
\mbox{Prob}[\xi^{\mu}_i = 1] = 1 - \mbox{Prob}[\xi^{\mu}_i = 0] = f. \label{eq.pattern}
\end{equation}
The expectation of $\bm{\xi}^{\mu}$ is $\mbox{E}[\xi^{\mu}_i] = f$;
thus, $f$ is considered to be the mean firing rate of
the memory pattern.
The memory pattern is sparse when $f \rightarrow 0$,
and this coding scheme is called sparse coding.

The synaptic weight $J_{ij}$ follows the form of synaptic
plasticity, which depends on the difference in spike times between 
the $i$-th (post-) and $j$-th (pre-) neurons.
This difference determines whether LTP or LTD is induced.
Such a learning rule is called spike-timing-dependent
plasticity (STDP).
Biological experimental findings have shown that LTP or LTD is
induced when the difference between the pre- and post-synaptic spike
times falls within about $20$ ms \cite[]{Zhang98} (Figure
\ref{fig.asymmetric}(a)).
There are also other types of STDP \cite[]{Abbott00}.
The STDP type of Figure \ref{fig.asymmetric}(a) has often been used in theoretical work (e.g., \cite{Song01, Rao00}).
We transformed this type of window function into the time window shown in Figure \ref{fig.asymmetric}(b).
This time window shows that LTP is induced when the $j$-th neuron fires one time step before the $i$-th neuron fires, i.e., $\xi_i^{\mu+1}\xi_j^{\mu}=1$, while LTD is induced when the $j$-th neuron fires one time step after the $i$-th neuron fires, i.e., $\xi_i^{\mu-1}\xi_j^{\mu}=1$.
We define a single time step in equations (\ref{eq.dynamics1}--\ref{eq.dynamics3}) as $20$ ms, and any duration of less than $20$ ms is ignored.
A synaptic plasticity obeying this time window is described by
\begin{equation}
\Delta J_{ij}= \xi_i^{\mu+1}\xi_j^{\mu} - (1+\epsilon)\xi_i^{\mu-1}\xi_j^{\mu}.
\end{equation}
After memory patterns are embedded by STDP, the synaptic weight is written as
\begin{equation}
J_{ij} = \frac{1}{N f (1 - f)} \sum_{\mu=1}^p ( \xi_i^{\mu + 1} \xi_j^{\mu} - ( 1 + \epsilon )\xi_i^{\mu - 1} \xi_j^{\mu}),
 \label{eq.learning}
\end{equation}
where the value $\frac{1}{Nf(1-f)}$ is a scaling factor and the number of memory patterns is $p= \alpha N$ where $\alpha$ is defined as the loading rate.
If both the $i$-th neuron and the $j$-th neuron fire simultaneously, neither LTP nor LTD occurs because equation (\ref{eq.learning}) does not include the term of $\xi_i^{\mu}\xi_j^{\mu}$.
Since the relative LTP and LTD magnitudes are more critical
than the absolute magnitudes, the LTD magnitude changes while
the LTP magnitude and the time duration are fixed. 
$\epsilon$ denotes deviation from a precise balance.
When $\epsilon=0$, the balance is precisely maintained and this
model is equivalent to the previous one \cite[]{Matsumoto02}.
A sequence of $p$ memory patterns is stored by STDP:
$\bm{\xi}^1 \rightarrow \bm{\xi}^2 \rightarrow \cdots
\rightarrow \bm{\xi}^p \rightarrow \bm{\xi}^1 \rightarrow
\cdots$.
In other words, $\bm{\xi}^1$ is retrieved at $t = 1$,
$\bm{\xi}^2$ is retrieved at $t = 2$, and $\bm{\xi}^{1}$ is
retrieved at $t = p+1$.
There is a critical value $\alpha_C$ of the loading rate, so that
a loading rate higher than $\alpha_C$ causes retrieval of the
pattern sequence to become unstable.
$\alpha_C$ is therefore called the storage capacity.

\section{Theory}
\subsection{The Case where Balance is Precisely Maintained}
\label{sec.previous}
Here, we briefly discuss the case where the balance is precisely maintained, which we previously discussed in detail \cite[]{Matsumoto02}.
This discussion will help explain the case where the balance is not precisely maintained.

First, we consider a simple situation where there are very few
memory patterns relative to the number of neurons; i.e., $p \sim
O(1)$. 
Let the state at time $t$ be equivalent to the $t$-th memory
pattern: $\bm{x}(t)=\bm{\xi}^t$.
The internal potential $u_i(t)$ of equation (\ref{eq.dynamics1})
is then given by
\begin{equation}
u_i(t) = \xi_i^{t+1} - \xi_i^{t-1}. \label{eq.u}
\end{equation}
$u_i(t)$ depends on two independent random variables, 
$\xi_i^{t+1}$ and $\xi_i^{t-1}$.
The first term $\xi_i^{t+1}$ is the signal term for the recall of
the pattern $\bm{\xi}^{t+1}$, which is intended to be retrieved at
time $t+1$, and the second term $\xi_i^{t-1}$ may interfere with
the retrieval of $\bm{\xi}^{t+1}$.
According to equation (\ref{eq.u}), $u_i(t)$ takes a value of
$0$, $-1$ or $+1$.
The probability distribution of $u_i(t)$ is given by Figure
\ref{fig.uprob}(a); i.e.,
\begin{equation}
\mbox{Prob}(u_i(t)) = (f-f^2) \delta(u_i(t)-1) + (1 - 2f + 2 f^2) \delta(u_i(t)) + (f-f^2) \delta(u_i(t)+1).
\end{equation}
If the threshold $\theta$ is set between $0$ and $+1$, the
probability of $u_i(t)=0$ is $1-2f+2f^2$, that of $u_i(t)=+1$
is $f-f^2$, and that of $u_i(t)=-1$ is $f-f^2$.
The overlap between the state $\bm{x}(t+1)$ and the memory
pattern $\bm{\xi}^{t+1}$ is given by
\begin{equation}
m^{t+1}(t+1) = \frac{1}{N f(1-f)} \sum_{i=1}^N ( \xi^{t+1}_i - f ) x_i(t+1) = 1 - f. \label{eq.overlap}
\end{equation}
At a sparse limit, the overlap $m^{t+1}(t+1)$ approaches $1$.
In other words, the memory pattern $\bm{\xi}^{t+1}$ is
retrievable.

Next, we consider the case where there is a large number of memory
patterns, i.e., $p \sim O(N)$.
The $i$-th neuronal internal potential $u_i(t)$ at time $t$ is
given using the periodic boundary condition of
$\xi_i^{p+1}=\xi_i^1$ and $\xi_i^{0}=\xi_i^p$ by
\begin{eqnarray}
u_i(t) \!\!\!\! &=& \!\!\!\! 
\frac{1}{N f(1-f)} \sum_{j=1}^N \sum_{\mu=1}^p (\xi_i^{\mu+1} \xi_j^{\mu} - \xi_i^{\mu-1} \xi_j^{\mu} ) x_j(t) \\
&=& \!\!\!\! ( \xi_i^{t+1} - \xi_i^{t-1} ) m^t(t) + z_i(t), \label{eq.u1}\\
z_i(t) \!\!\!\! &=& \!\!\!\! \sum_{\mu \neq t}^p (\xi_i^{\mu+1} - \xi_i^{\mu-1}) m^{\mu}(t).
\end{eqnarray}
The first term in equation (\ref{eq.u1}) is the signal term for
the recall of pattern $\bm{\xi}^{t+1}$.
The second term is the cross-talk noise term that represents the contributions of non-target patterns other than $\bm{\xi}^{t-1}$, and prevents the retrieval of the target pattern $\bm{\xi}^{t+1}$.
The cross-talk noise is assumed to obey a Gaussian distribution
with mean $0$ and time-dependent variance $\sigma^2(t)$.
We derive the recursive equation for the overlap $m^t(t)$ between the state $\bm{x}(t)$ and the target pattern $\bm{\xi}^{t}$ by using the method of the statistical neurodynamics \cite[]{Amari88,Okada95,Matsumoto02} to calculate the overlap $m^t(t)$.
The recursive equation is given by
\begin{equation}
m^t(t) = \frac{ 1 - 2f }{2} \mbox{erf}( \phi_0 ) - \frac{ 1 - f }{2} \mbox{erf}( \phi_1 ) + \frac{f}{2} \mbox{erf}( \phi_2 ),
\end{equation}
where $\phi_0 = \frac{\theta}{\sqrt{2} \sigma(t-1)}, \phi_1 = \frac{-m^{t-1}(t-1)+\theta}{\sqrt{2} \sigma(t-1)}, \phi_2 = \frac{m^{t-1}(t-1)+\theta}{\sqrt{2} \sigma(t-1)}$.
Since $\phi_0$, $\phi_1$, and $\phi_2$ include $\sigma(t-1)$, $m^t(t)$ depends on $\sigma(t)$.
The recursive equation for $\sigma(t)$ is
\begin{equation}
\sigma^2(t) = \sum_{a=0}^t {}_{2(a+1)}\mbox{C}_{(a+1)} \alpha q(t-a) \prod_{b=1}^a U^2(t-b+1).
\end{equation}
The recursive equations for $U(t)$ and $q(t)$ are given by
\begin{equation}
U(t) = \frac{1}{\sqrt{2 \pi} \sigma(t-1)} \{ ( 1 - 2f + 2f^2 ) e^{-\phi_0^2} + f( 1 - f )( e^{- \phi_1^2} + e^{- \phi_2^2} ) \},
\end{equation}
and
\begin{equation}
q(t) = \frac{1}{2} \left( 1 - ( 1 - 2f + 2f^2) \mbox{erf}(\phi_0) - f(1-f) ( \mbox{erf}(\phi_1) + \mbox{erf}(\phi_2) ) \right), \label{eq.q}
\end{equation}
where $\mbox{erf}(y) = \frac{2}{\sqrt{\pi}} \int_0^y \exp{(-u^2)} du$, ${}_{b}\mbox{C}_{a} = \frac{b!}{a! (b-a)!}$, $a! = a \times (a-1) \times \cdots \times 1$.
The details of this derivation are given in the Appendix.
To obtain the storage capacity $\alpha_C$, we calculate the overlap in the steady state.
We set the initial state of the network at the first memory pattern: $\bm{x}(1)=\bm{\xi}^1$.
Setting the initial values at $m^1(1)=1$, $\sigma^2(1)=2\alpha f$, $U(1)=0$, and $q(1)=f$ and using the recursive equations (13-16), we obtain the overlap in the steady state.
There is a critical value $\alpha_C$ of the loading rate, so that the loading rate higher than $\alpha_C$ causes retrieval of the pattern sequence to become unstable.
In computer simulation, when the overlap in the steady state is smaller than $0.5$, the critical loading rate is regarded as the storage capacity $\alpha_C$.

\subsection{The Case Where Balance is not Precisely Maintained}
The case where balance is not precisely maintained is difficult to treat because the mean of the cross-talk noise is not $0$.
However, previous work using computer simulations has shown that
the stored limit cycle is stably retrieved when the number of
neurons is finite \cite[]{Munro00,Yoshioka02}.
However, this mathematical mechanism has remained unknown.
To investigate this unknown mechanism, we will use the theory derived
in section \ref{sec.previous}.

First, let the state at time $t$ be equivalent to the $t$-th
memory pattern: $\bm{x}(t)=\bm{\xi}^t$.
The $i$-th neuronal internal potential $u_i(t)$ at time $t$
is then given using the periodic boundary condition of
$\xi_i^{p+1}=\xi_i^1$ and $\xi_i^{0}=\xi_i^p$ by 
\begin{eqnarray}
u_i(t) \!\!\!\! &=& \!\!\!\!
\frac{1}{N f(1-f)} \sum_{j=1}^N \sum_{\mu=1}^p (\xi_i^{\mu+1}\xi_j^{\mu} - (1+\epsilon)\xi_i^{\mu-1}\xi_j^{\mu}) x_j(t) \\
\!\!\!\! &=& \!\!\!\! \frac{1}{N f(1-f)} \sum_{j=1}^N \sum_{\mu=1}^p (\bar{\xi}_i^{\mu+1} - \bar{\xi}_i^{\mu-1}) \bar{\xi}_j^{\mu} x_j(t) \nonumber \\
& & \!\!\!\! - \frac{\epsilon}{N f(1-f)} \sum_{j=1}^N \sum_{\mu=1}^p \xi_i^{\mu-1}\xi_j^{\mu} x_j(t), \label{eq.epsilonu}
\end{eqnarray}
where $\bar{\xi}_j^{\mu}=\xi_j^{\mu}-f$.
Here, we evaluate the second term; that is, the compensation term in
equation (\ref{eq.epsilonu}).
The average of this term is $\frac{\alpha f}{1-f} \epsilon N q(t)$ 
where $q(t)=\frac{1}{N} \sum_{j=1}^N x_j(t)=f$ when
$\bm{x}(t)=\bm{\xi}^t$.
A typical value of this term is $0.19$ for $N=5000$,
$\alpha=0.067$, $\epsilon=0.05$ and $f=0.1$, which are typical
values used in the following analysis.
This value is small enough for the overlap $m^{t}(t) \simeq 1$.
The fluctuation in the compensation term can be eliminated
since the order of the variance of this term is smaller than that of
the average.
Therefore, the average term of the compensation term adds to the
signal term in equation (\ref{eq.u1}).

Next, we will discuss a mechanism that enables memory patterns to be
retrieved in the small $\epsilon$ and finite $N$ case.
The internal potential with $\bm{x}(t)=\bm{\xi}^t$ is given by
\begin{equation}
u_i(t) = \xi_i^{t+1} - \xi_i^{t-1} - \frac{\alpha f^2}{1-f} \epsilon N.
\end{equation}
When $\epsilon \neq 0$, each distribution of $u_i(t)$ shifts by
$\frac{\alpha f^2}{1-f} \epsilon N$ as shown in Figure
\ref{fig.uprob}(b).
For the typical case at $N=5000$, $\alpha=0.067$,
$\epsilon=0.05$, $f=0.1$, and $\theta=0.52$, each
distribution of $u_i(t)$ would shift left by $0.19$.
Since the threshold is $0.52$, the probability of
$u_i(t+1)=+1$ is little changed.
In other words, for the small $\epsilon$ and finite $N$ case, retrieval of the next pattern $\bm{\xi}^{t+1}$ is successful.
Otherwise, the retrieval fails.

Since the average term of the compensation term adds to the
signal term in equation (\ref{eq.u1}), the $i$-th neuronal
internal potential $u_i(t)$ at time $t$ is given by
\begin{equation}
u_i(t) =
( \bar{\xi}_i^{t+1} - \bar{\xi}_i^{t-1} ) m^{t}(t) + \sum_{\mu \neq t}^{p} ( \bar{\xi}_i^{\mu+1} - \bar{\xi}_i^{\mu-1} ) m^{\mu}(t) - \frac{\alpha \epsilon f N}{1-f}q(t).
\end{equation}
Here, the first term is the signal term and the second is the cross-talk
noise term.
The third term is the compensation term.
Statistical neurodynamics enables us to derive dynamical equations.
These equations are almost the same as those in section \ref{sec.previous}
other than $\phi_0$, $\phi_1$ and $\phi_2$.
In $\phi_0$, $\phi_1$ and $\phi_2$, $\theta$ is replaced by
$\theta + \frac{\alpha \epsilon f N}{1-f}q(t-1)$.

\section{Results}
We used statistical neurodynamics and computer simulation to
investigate the properties of our model and examine its behavior.

Figure \ref{fig.alpha} shows the dependence of the overlap $m^t(t)$
on the loading rate $\alpha$ when the mean firing rate of the
memory pattern is $f=0.1$ and the threshold (optimized to maximize the storage capacity) is $\theta=0.52$.
The solid line denotes the steady-state values of the overlap
$m^t(t)$ at $\epsilon=0.0$ in (a) and $\epsilon=0.05$ in (b).
We obtained $m^t(t)$ by setting the initial state of the network
at the first memory pattern: $\bm{x}(1)=\bm{\xi}^1$.
The storage capacity is $\alpha_C=0.27$ in (a) and
$\alpha_C=0.067$ in (b).
The data points and error bars (with the former indicating median values and the latter indicating 1/4 and 3/4 deviations) show the computer simulation results from $11$ trials with $5000$ neurons ($N=5000$).
The discrepancy between the values of $m_C$ obtained from the computer
simulations and the analytical results was caused by the
finite size effect of the computer simulations \cite[]{Matsumoto02}.

Figure \ref{fig.epsilondepend}(a) shows the storage capacity
$\alpha_C$ as a function of $\epsilon$.
In the computer simulaton, when the overlap in the steady state is smaller than $0.5$,
the critical loading rate $\alpha$ is regarded as the storage
capacity $\alpha_C$.
The data points and error bars (with the former indicating mean
values and the latter indicating standard deviations) show the 
computer simulation results from $10$ trials at $f=0.1$, $\theta=0.52$, and $N=3000$
($\square$) or $N=5000$ ($\circ$).
The solid line shows the theoretical results at $N=3000$, while
the dashed line and the gray solid line show the results at
$N=5000$ and $N=100000$, respectively.
As $|\epsilon|$ increased, $\alpha_C$ decreased.
When $\epsilon=0.5$, the storage capacities for $N=3000$ and
$N=5000$ were $0.017$ and $0.011$, respectively.
However, the storage capacity for $N=100000$ was $0$.
In other words, when the number of neurons is finite, $\alpha_C$
takes a finite value.
These computer simulation results are consistent with the theoretical
results at small $\epsilon$.
Figure \ref{fig.epsilondepend}(b) shows the maximum number of storable memory patterns, $p_{max}=\alpha_C N$, as a function of $\epsilon$.
This is replot of Figure \ref{fig.epsilondepend}(a).
The solid line shows the theoretical results at $N=3000$, while
the dashed line and the gray solid line show the results at
$N=5000$ and $N=100000$, respectively.
As $|\epsilon|$ increased, actual number of storable memory patterns decreased.

Figure \ref{fig.limitepsilon} shows $\alpha_C$ as a function of
$\epsilon$.
The solid line shows the theoretical results at $f=0.1$ and
$\theta=0.52$, while the dashed line shows
$\log_{10}{\alpha_C}=-\log_{10}{N \epsilon}+1.435$
at $N=5000$, $f=0.1$, and $\theta=0.52$. 
This figure shows that $\alpha_C$ converged to $0$ as the order of $\frac{1}{N \epsilon}$ (i.e., $O(\frac{1}{N \epsilon})$) at a large $N \epsilon$ limit.

\section{Summary and Discussion}
We have investigated, using an associative memory network, how the balance between the LTP and LTD of STDP affects the retrieval of spatio-temporal patterns.
Other authors have \textit{numerically} investigated the impact of an
imbalance between LTP and LTD on network properties.
We have \textit{analytically} investigated the retrieval properties by applying statistical neurodynamics.
In the case where the value of LTD deviates from a precise balance the stored limit cycle using STDP is unstable at the thermodynamic limit.
However, the stored limit cycle is stable when the number of neurons is \textit{finite}.
Moreover, computer simulation results were consistent with the theoretical results at small $\epsilon$.
Thus, a network containing a \textit{finite} number of neurons
becomes robust against deviation from a precise LTP/LTD balance.
Since the number of neurons in the brain is considered to be
finite, our results might be applicable to the brain.
Furthermore, the storage capacity converges to $0$ as
$O(\frac{1}{N \epsilon})$ at a large $N \epsilon$ limit.

\clearpage
\appendix
\section{Derivation of Recursive Equations when Balance is Precisely Maintained}
\label{sec.appendix}
In this appendix, we derive recursive equations for the overlap 
$m^t(t)$ when the balance between the LTP and LTD of STDP is precisely maintained.
We divide an internal potential $u_i(t)$ at time $t$ into two parts, a signal term for a retrieval of a target pattern and a cross-talk noise term that represents the contributions from non-target patterns and prevents the target pattern from being retrieved.
$u_i(t)$ is derived from equation (\ref{eq.learning}) and the periodic boundary conditions
of $\xi_i^{p+1}=\xi_i^1$ and $\xi_i^{0}=\xi_i^{p}$ in the
following way:
\begin{eqnarray}
u_i(t) \!\!\!\! &=& \!\!\!\! \frac{1}{N f(1-f)} \sum_{j=1}^N \sum_{\mu=1}^p (\xi_i^{\mu+1} \xi_j^{\mu} - \xi_i^{\mu-1} \xi_j^{\mu} ) x_j(t) \\
&=& \!\!\!\! ( \xi_i^{t+1} - \xi_i^{t-1} ) m^t(t) + z_i(t),
\end{eqnarray}
where $z_i(t)$ is given by
\begin{equation}
z_i(t) = \sum_{\mu \neq t}^p (\xi_i^{\mu+1} - \xi_i^{\mu-1}) m^{\mu}(t),
\end{equation}
and $m^{\mu}(t)$ is the overlap between $\bm{\xi}^{\mu}$ and $\bm{x}(t)$ and which is given by
\begin{equation}
m^{\mu}(t) = \frac{1}{Nf(1-f)} \sum_{i=1}^N ( \xi_i^{\mu} - f ) x_i(t).
\end{equation}
The first term in equation (22) is the signal term and the second term is the cross-talk noise term.
Since $x_i(t)$ in equation (24) depends on $\xi_i^{\mu}$, the distribution of the cross-talk noise term $z_i(t)$ is unknown.
The dependence on $\xi_i^{\mu}$ is extracted from $x_i(t)$ using the Taylor expansion.

At time $t$, the pattern $\bm{\xi}^t$ is designed to be retrieved.
Therefore, we can assume that $m^t(t)$ is order of 1 with respect to $N$ (i.e., $m^t(t) \sim O(1)$) and $m^{\mu}(t)$($\mu \neq t$) is order of $1/\sqrt{N}$ with respect to $N$ (i.e., $m^{\mu}(t) \sim O(1/\sqrt{N})$).
Since $m^{\mu}(t) \sim O(1/\sqrt{N})$, $m^{\mu-1}$ and $m^{\mu+1}$ are order of $1/\sqrt{N}$ with respect to $N$.
At a thermodynamic limit, $N \rightarrow \infty$, $m^{\mu-1}$ and $m^{\mu+1}$ are small.
To extract the dependence on $\xi^{\mu}$ from $x_i(t)$, the $i$-th neuronal state at time $t+1$ is transformed:
\begin{eqnarray}
x_i(t+1) \!\!\!\! &=& \!\!\!\! F( \sum_{\nu=1}^{p} ( \xi_i^{\nu + 1} - \xi_i^{\nu - 1} ) m^{\nu}(t) - \theta ) \\
&=& \!\!\!\! F( \sum_{\nu \neq \mu}^{p} ( \bar{\xi}_i^{\nu + 1} - \bar{\xi}_i^{\nu - 1} ) m^{\nu}(t) - \theta ) \nonumber \\
& & \!\!\!\! + \bar{\xi}_i^{\mu}(m^{\mu-1}(t) - m^{\mu+1}(t)) F'( \sum_{\nu \neq \mu}^{p} ( \bar{\xi}_i^{\nu + 1} - \bar{\xi}_i^{\nu - 1} ) m^{\nu}(t) - \theta ) \\
&=& \!\!\!\! x_i^{(\mu)}(t+1) + \bar{\xi}_i^{\mu}(m^{\mu-1}(t) - m^{\mu+1}(t)) x_i'^{(\mu)}(t+1), 
\end{eqnarray}
where $\bar{\xi}_i^{\mu}=\xi_i^{\mu} - f $, $x_i^{(\mu)}(t+1)$ does not include $\xi_i^{\mu}$, and $x_i'^{(\mu)}(t)$ is the differential of $x_i^{(\mu)}(t)$.
Using this relationship, we derive the cross-talk noise at time $t$:
\begin{eqnarray}
z_i(t) \!\!\!\! &=& \!\!\!\! \sum_{\mu \neq t}^{p} (\xi_i^{\mu+1} - \xi_i^{\mu-1}) m^{\mu}(t) \\
&=& \!\!\!\! \frac{1}{N f(1-f)} \sum_{\mu \neq t}^{p} \sum_{j=1}^N
( \bar{\xi}_i^{\mu+1} - \bar{\xi}_i^{\mu-1} ) \bar{\xi}_j^{\mu}
x_j^{(\mu)}(t) \nonumber \\
& & \!\!\!\! + \sum_{\nu \neq t}^{p} U(t) ( \bar{\xi}_i^{\nu+1} m^{\nu-1}(t-1) - 2 \bar{\xi}_i^{\nu-1} m^{\nu-1}(t-1) + \bar{\xi}_i^{\nu-1} m^{\nu+1}(t-1) ),
\end{eqnarray}
where $U(t) = \mbox{E} [ x_i'^{(\mu)}(t) ]$. 
Therefore, the square of $z_i(t)$ is given by
\begin{eqnarray}
(z_i(t))^2 \!\!\!\! &=& \!\!\!\! \left( \frac{1}{N f(1-f)} \right)^2 \sum_{\mu \neq t}^{p} \sum_{j=1}^N ( \bar{\xi}_i^{\mu+1} - \bar{\xi}_i^{\mu-1} )^2 ( \bar{\xi}_j^{\mu} )^2 ( x_j^{(\mu)}(t) )^2 \nonumber \\
& & \!\!\!\! + \sum_{\nu \neq t}^{p} U(t)^2 ( \bar{\xi}_i^{\nu+1} m^{\nu-1}(t-1) - 2 \bar{\xi}_i^{\nu-1} m^{\nu-1}(t-1) \nonumber \\
& & \!\!\!\! + \bar{\xi}_i^{\nu-1} m^{\nu+1}(t-1) )^2 \\
&=& \!\!\!\! ( 1^2 + (-1)^2 ) \alpha q(t) + ( 1^2 + (-2)^2 + 1^2 ) \alpha q(t-1) U^2(t) \nonumber \\
& & \!\!\!\! + ( 1^2 + (-3)^2 + 3^2 + (-1)^2 ) \alpha q(t-2) U^2(t) U^2(t-1) + \cdots \\
&=& \!\!\!\! \sum_{a=0}^t {}_{2(a+1)}\mbox{C}_{(a+1)} \alpha q(t-a)
\prod_{b=1}^a U^2(t-b+1),
\end{eqnarray}
where $p = \alpha N$, $q(t) = \frac{1}{N} \sum_{i=1}^N
(x_i^{(\mu)}(t))^2$, ${}_{b}\mbox{C}_{a} = \frac{b!}{a! (b-a)!}$
and $a!$ is a factorial with positive integer $a$.
We applied the relationship $\sum_{a=0}^b({}_{b}\mbox{C}_{a})^2 =
{}_{2b}\mbox{C}_{b}$ in this derivation.
At a thermodynamic limit, $N \rightarrow \infty$, $m^{\mu}(t)$ tends to be deterministic.
Therefore, $x_i^{(\mu)}(t)$ is independent of $\xi_i^{\mu}$.
This enables us to assume that the cross-talk noise term obeys a Gaussian distribution with a mean of $0$ and time-dependent variance $\sigma^2(t)$: $\mbox{E}[ z_i(t) ] =0, \mbox{E}[ (z_i(t))^2 ] = \sigma^2(t)$.
Since the distribution of the cross-talk noise term is known, recursive equations for $m^t(t)$ and $\sigma^2(t)$ are obtained.
We get the recursive equation for $\sigma^2(t)$:
\begin{equation}
\sigma^2(t) = \sum_{a=0}^t {}_{2(a+1)}\mbox{C}_{(a+1)} \alpha q(t-a)
\prod_{b=1}^a U^2(t-b+1).
\end{equation}
The overlap between the state $\bm{x}(t)$ and the target pattern $\bm{\xi}^t$ is given by
\begin{eqnarray}
m^t(t) \!\!\!\! &=& \!\!\!\! \frac{1}{Nf(1-f)} \sum_{i=1}^N ( \xi_i^t - f ) x_i(t) 
= \frac{1}{Nf(1-f)} \sum_{i=1}^N ( \xi_i^t - f ) F(u_i(t)) \\
&=& \!\!\!\! \frac{1}{Nf(1-f)} \sum_{i=1}^N ( \xi_i^t - f ) F((\xi_i^t - \xi_i^{t-2}) m^{t-1}(t-1) + z_i(t-1) - \theta).
\end{eqnarray}
Since $u_i(t)$ is independent and identical distribution (i.i.d.), by the law of large numbers, we can replace the average over $i$ by an average over the memory patterns $\xi^{\mu}$ and the Gaussian noise term $z \sim \mathcal{N}(0,\sigma^2)$.
Then, the recursive equation for the overlap $m^t(t)$ is transformed:
\begin{eqnarray}
m^t(t) \!\!\!\! &=& \!\!\!\! \frac{1}{f(1-f)} \frac{1}{\sqrt{2 \pi}\sigma} \int_{-\infty}^{\infty} dz e^{-\frac{z^2}{2 \sigma^2}} \langle \langle ( \xi^t - f ) \nonumber \\
& & \times F((\xi^t - \xi^{t-2}) m^{t-1}(t-1) + z - \theta) \rangle \rangle \\
&=& \!\!\!\! \frac{1}{f(1-f)} \frac{1}{\sqrt{2 \pi}} \int_{-\infty}^{\infty} d\tilde{z} e^{-\frac{z^2}{2}} \langle \langle ( \xi^t - f ) \nonumber \\
& & \times F((\xi^t - \xi^{t-2}) m^{t-1}(t-1) + \sigma(t-1) \tilde{z} - \theta) \rangle \rangle \\
&=& \!\!\!\! \frac{ 1 - 2f }{2} \mbox{erf}( \phi_0 ) - \frac{ 1 - f }{2} \mbox{erf}( \phi_1 ) + \frac{f}{2} \mbox{erf}( \phi_2 ),
\end{eqnarray}
where $\mbox{erf}(y) = \frac{2}{\sqrt{\pi}} \int_0^y \exp{(-u^2)}
du$, $\phi_0 = \frac{\theta}{\sqrt{2}\sigma(t-1)}$,
$\phi_1 = \frac{-m^{t-1}(t-1)+\theta}{\sqrt{2} \sigma(t-1)}$,
$\phi_2 = \frac{m^{t-1}(t-1)+\theta}{\sqrt{2} \sigma(t-1)}$, 
and $\langle \langle \cdot \rangle \rangle$ denotes an average over the memory pattern $\xi^{\mu}$.
Since $x_i'(t)-x_i'^{(\mu)}(t) \sim O(\frac{1}{\sqrt{N}})$ and the thermodynamic limit $N \rightarrow \infty$ is considered, $x_i'^{(\mu)}(t)=x_i'(t)$.
Using this relationship, we derive $U(t)$:
\begin{eqnarray}
U(t) \!\!\!\! &=& \!\!\!\! \frac{1}{N} \sum_{i=1}^N x_i'^{(\mu)}(t) = \frac{1}{N} \sum_{i=1}^N x_i'(t) \\
&=& \!\!\!\! \frac{1}{\sqrt{2 \pi}} \int_{-\infty}^{\infty} dz e^{-\frac{z^2}{2}}  \langle \langle F'( ( \xi^t - \xi^{t-2} ) m^{t-1}(t-1) + \sigma(t-1) z - \theta ) \rangle \rangle \\
&=& \!\!\!\! \frac{1}{\sqrt{2 \pi} \sigma(t-1)} \{ ( 1 - 2f + 2f^2 ) e^{- \phi_0^2} + f( 1 - f )( e^{- \phi_1^2} + e^{- \phi_2^2} ) \}.
\end{eqnarray}
Since $x_i(t)-x_i^{(\mu)}(t) \sim O(\frac{1}{\sqrt{N}})$ and $N \rightarrow \infty$, $x_i^{(\mu)}(t)=x_i(t)$.
Using this relationship, we derive $q(t)$:
\begin{eqnarray}
q(t) \!\!\!\! &=& \!\!\!\! \frac{1}{N} \sum_{i=1}^N (x_i^{(\mu)}(t))^2 = \frac{1}{N} \sum_{i=1}^N (x_i(t))^2 \\
&=& \!\!\!\!  \frac{1}{\sqrt{2 \pi}} \int_{-\infty}^{\infty} dz e^{-\frac{z^2}{2}}  \langle \langle F^2( ( \xi^t - \xi^{t-2} ) m^{t-1}(t-1) + \sigma(t-1) z - \theta ) \rangle \rangle \\
&=& \!\!\!\! \frac{1}{2} \left( 1 - ( 1 - 2f + 2f^2 ) \mbox{erf}(\phi_0) - f(1-f) ( \mbox{erf}(\phi_1) + \mbox{erf}(\phi_2) ) \right).
\end{eqnarray}

\clearpage
\pagestyle{empty}

\section*{Figure legends}
\begin{figure}[htb]
\caption{\label{fig.asymmetric}
The time window of spike-timing-dependent plasticity. 
(a): Results of the biological experiment
\cite[]{Zhang98}. (b): STDP in our model. LTP occurs when the
$j$-th neuron fires one time step before the $i$-th neuron. LTD
occurs when the $j$-th neuron fires one time step after the $i$-th
neuron.}
\end{figure}

\begin{figure}[htb]
\caption{\label{fig.uprob}
Probability distribution of the $i$-th neuronal
potential $u_i(t)$ at time $t$. (a)The balance between the LTP and
LTD of STDP is precisely maintained; i.e., $\epsilon=0$. (b)The
balance is not precisely maintained; i.e., $\epsilon \neq 0$. Each
distribution shifts by $\frac{\alpha f^2}{1-f} \epsilon N$.}
\end{figure}

\begin{figure}[htb]
\caption{\label{fig.alpha}
The solid line shows the overlap in the steady state when the
firing rate of stored patterns is $0.1$ at $\epsilon=0.0$ in (a)
and $\epsilon=0.05$ in (b). 
The storage capacity is $0.27$ in (a) and $0.067$ in (b).
In both figures, the threshold is $0.52$ and the number of neurons
is $5000$. 
The data points indicate the median values and the ends of the
error bars indicate 1/4 and 3/4 derivations, respectively, from $11$ computer simulation trials. }
\end{figure}

\clearpage
\begin{figure}[htb]
\caption{\label{fig.epsilondepend}
(a)The storage capacity $\alpha_C$ and (b)the maximum number of storable memory patterns, $p_{max}=\alpha_C N$, (b) as a function of $\epsilon$.
The data points and error bars (with the former indicating the mean values and latter
indicating standard deviations) in (a) show computer simulation results from $10$ trials at $N=3000$ ($\square$) or $N=5000$
($\circ$).
The solid line, dashed line and gray solid line show
the theoretical results at $N=3000$, $N=5000$, and $N=100000$,
respectively. All results were obtained at $f=0.1$ and
$\theta=0.52$. As $|\epsilon|$ increased, $\alpha_C$ and $p_{max}$ decreased.} 
\end{figure}

\begin{figure}[htb]
\caption{\label{fig.limitepsilon}
Storage capacity $\alpha_C$ as a function of $\epsilon$ at
$N=5000$, $f=0.1$, and $\theta=0.52$. The solid line shows the
theoretical results while the dashed line shows
$\log_{10}{\alpha_C}=-\log_{10}{N \epsilon}+1.435$.
$\alpha_C$ converged to $0$ as $O(\frac{1}{N \epsilon})$ at a
large $N \epsilon$ limit.}
\end{figure}

\begin{figure}[p]
\begin{center}
\Huge (a)\includegraphics[width=10.795cm]{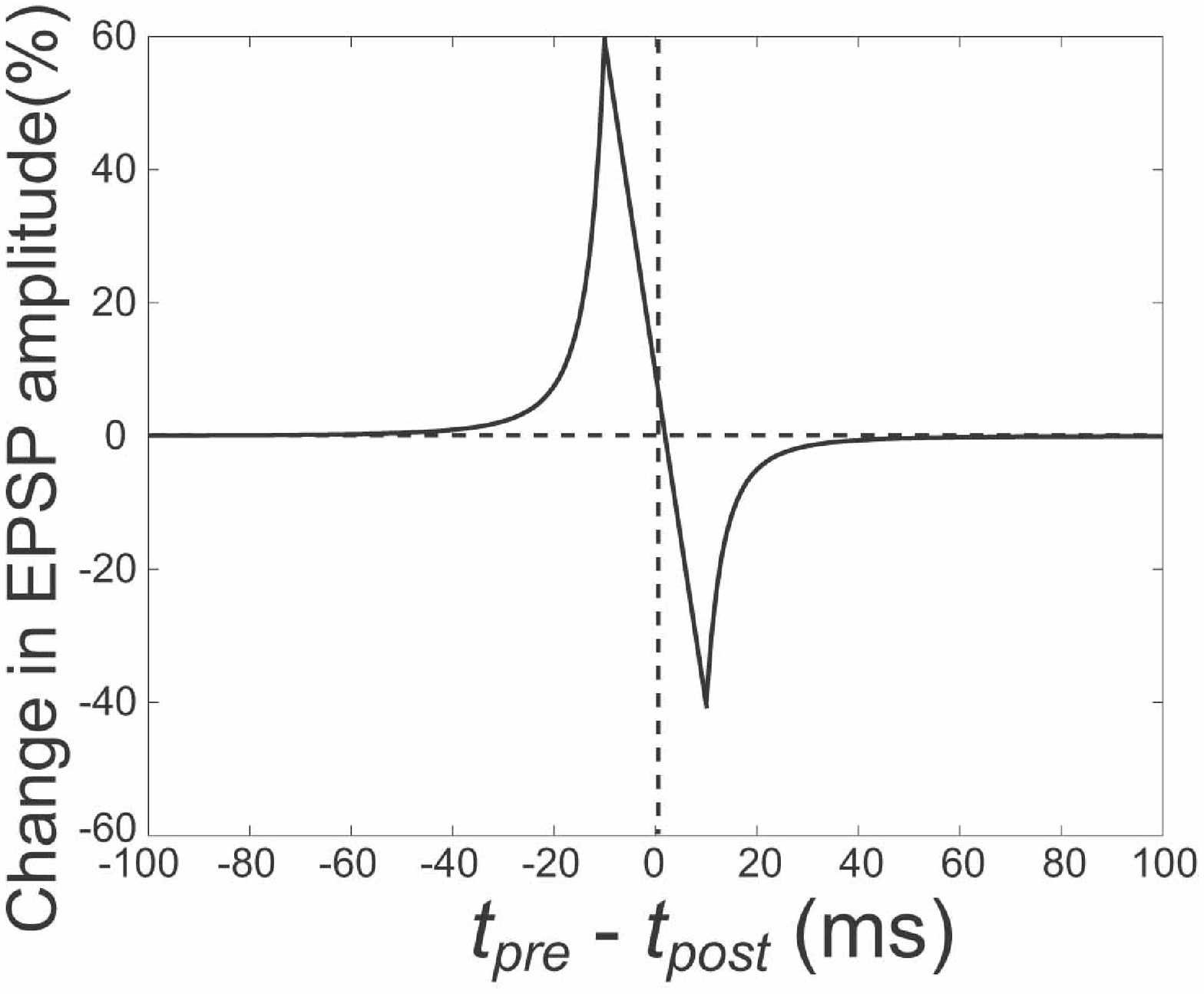}\\
\Huge (b)\includegraphics[width=10.795cm]{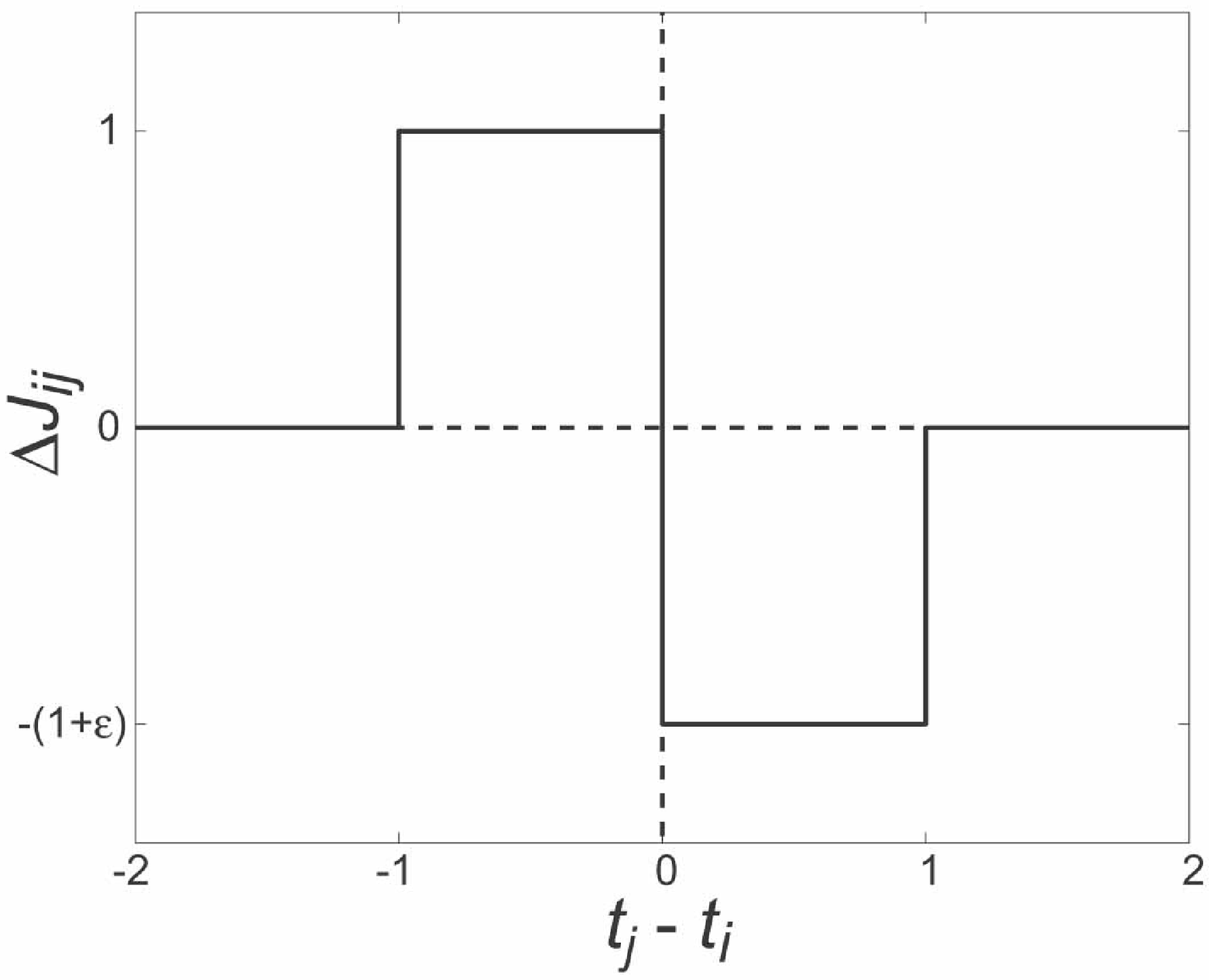}\\
\vspace{1.5cm}
{\Huge Figure 1}
\end{center}
\end{figure}

\clearpage
\begin{figure}[p]
\begin{center}
\Huge (a)\includegraphics[width=10.795cm]{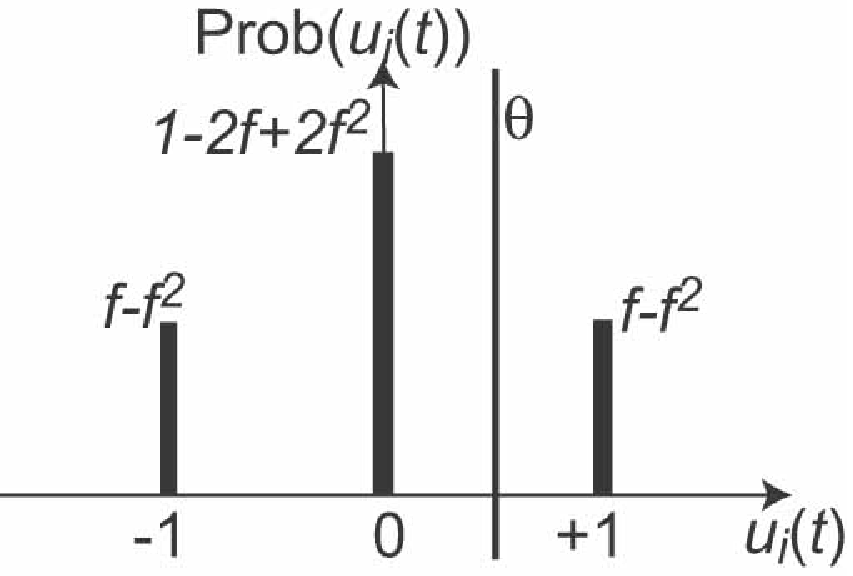}\\
\Huge (b)\includegraphics[width=10.795cm]{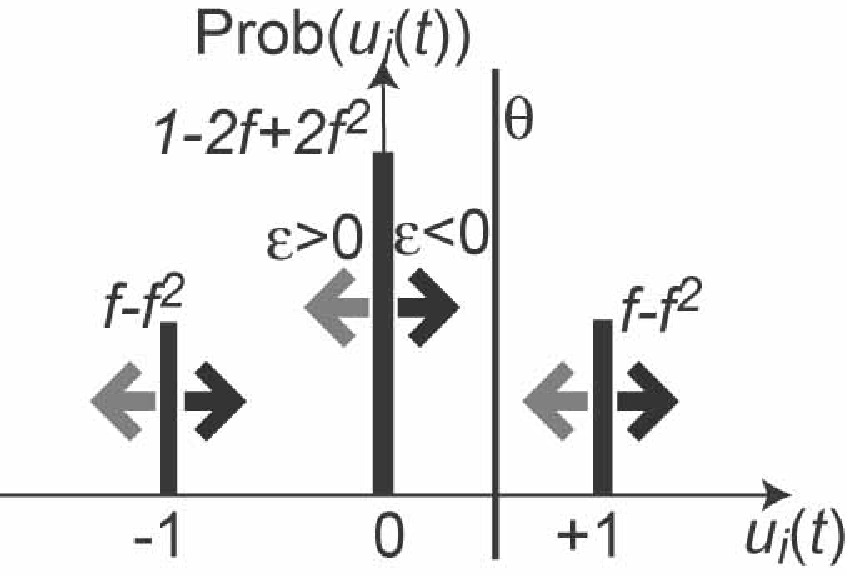}\\
\vspace{1.5cm}
{\Huge Figure 2}
\end{center}
\end{figure}

\clearpage
\begin{figure}[p]
\begin{center}
\Huge (a)\includegraphics[width=10.795cm]{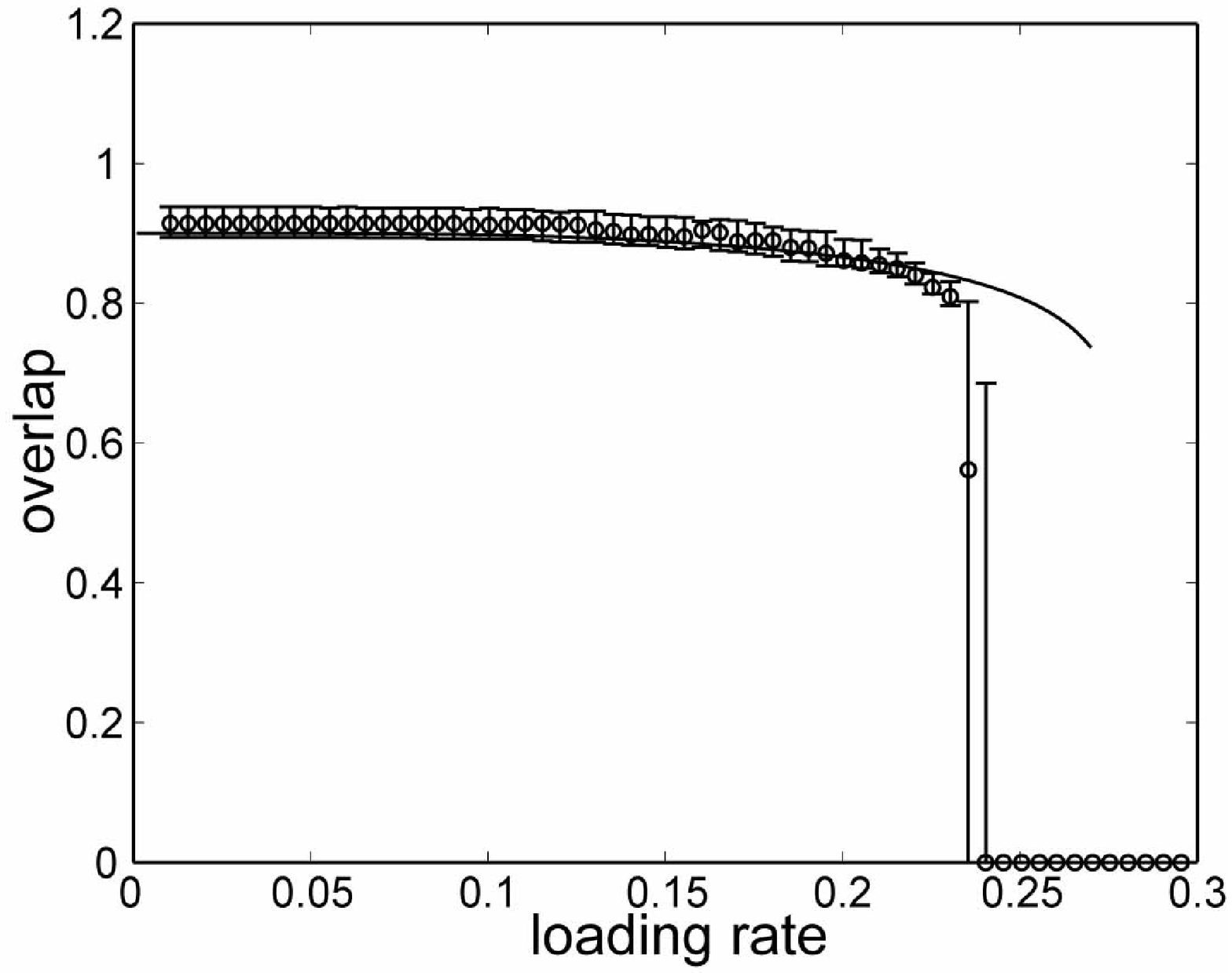}\\
\Huge (b)\includegraphics[width=10.795cm]{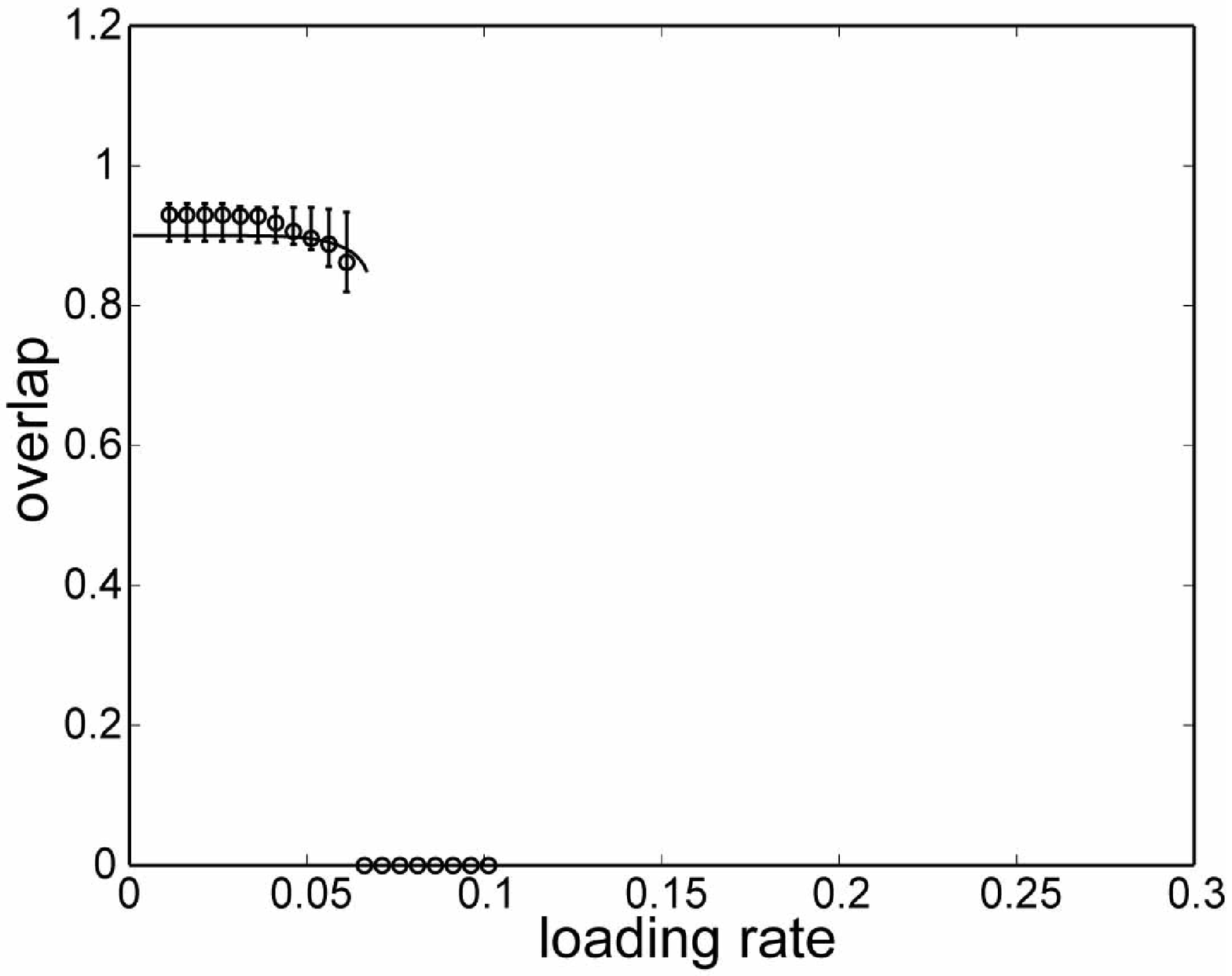}\\
\vspace{1.5cm}
{\Huge Figure 3}
\end{center}
\end{figure}

\clearpage
\begin{figure}[p]
\begin{center}
\Huge (a)\includegraphics[width=10.795cm]{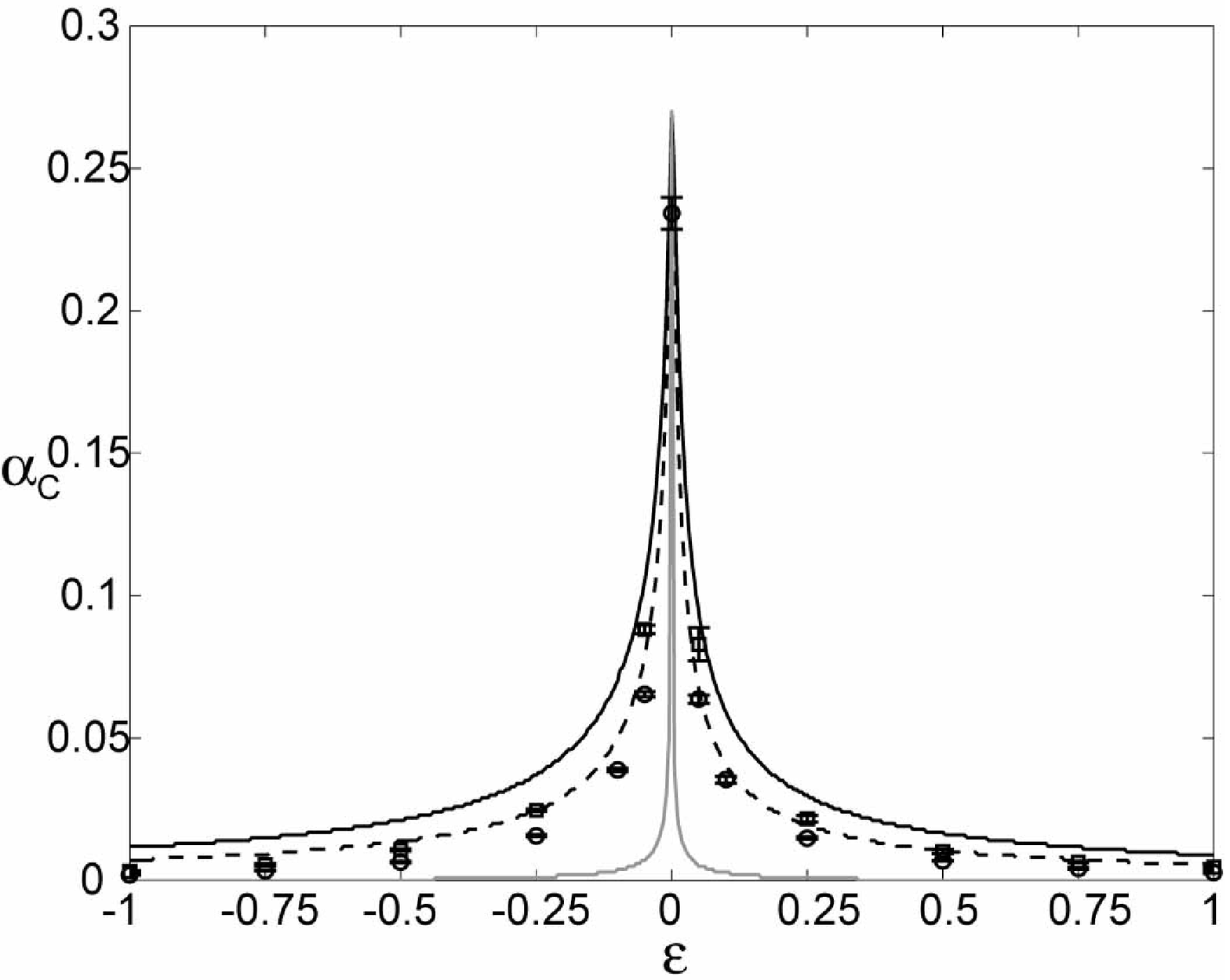}\\
\Huge (b)\includegraphics[width=10.795cm]{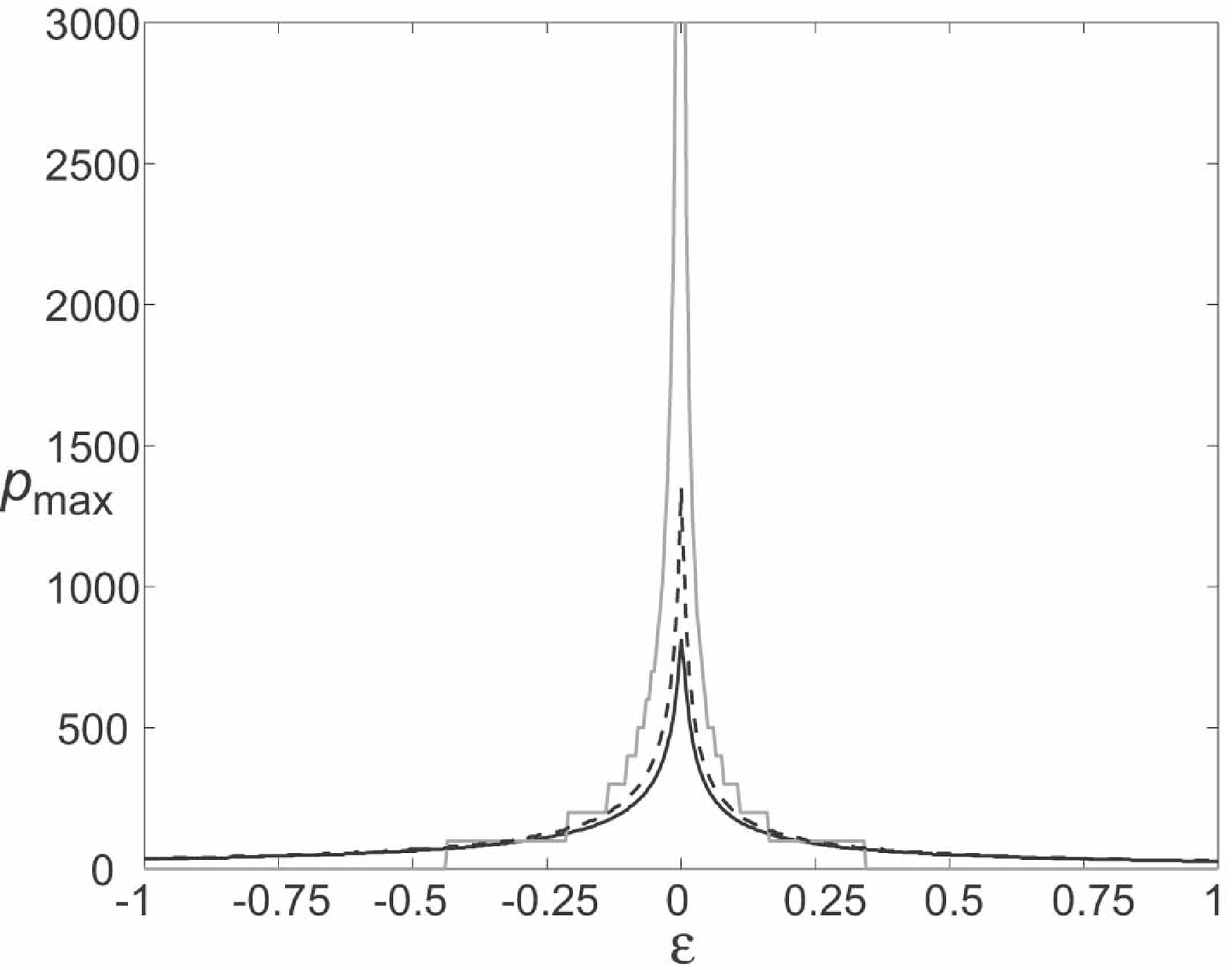}\\
\vspace{1.5cm}
{\Huge Figure 4}
\end{center}
\end{figure}

\clearpage
\begin{figure}[p]
\begin{center}
\includegraphics[width=10.795cm]{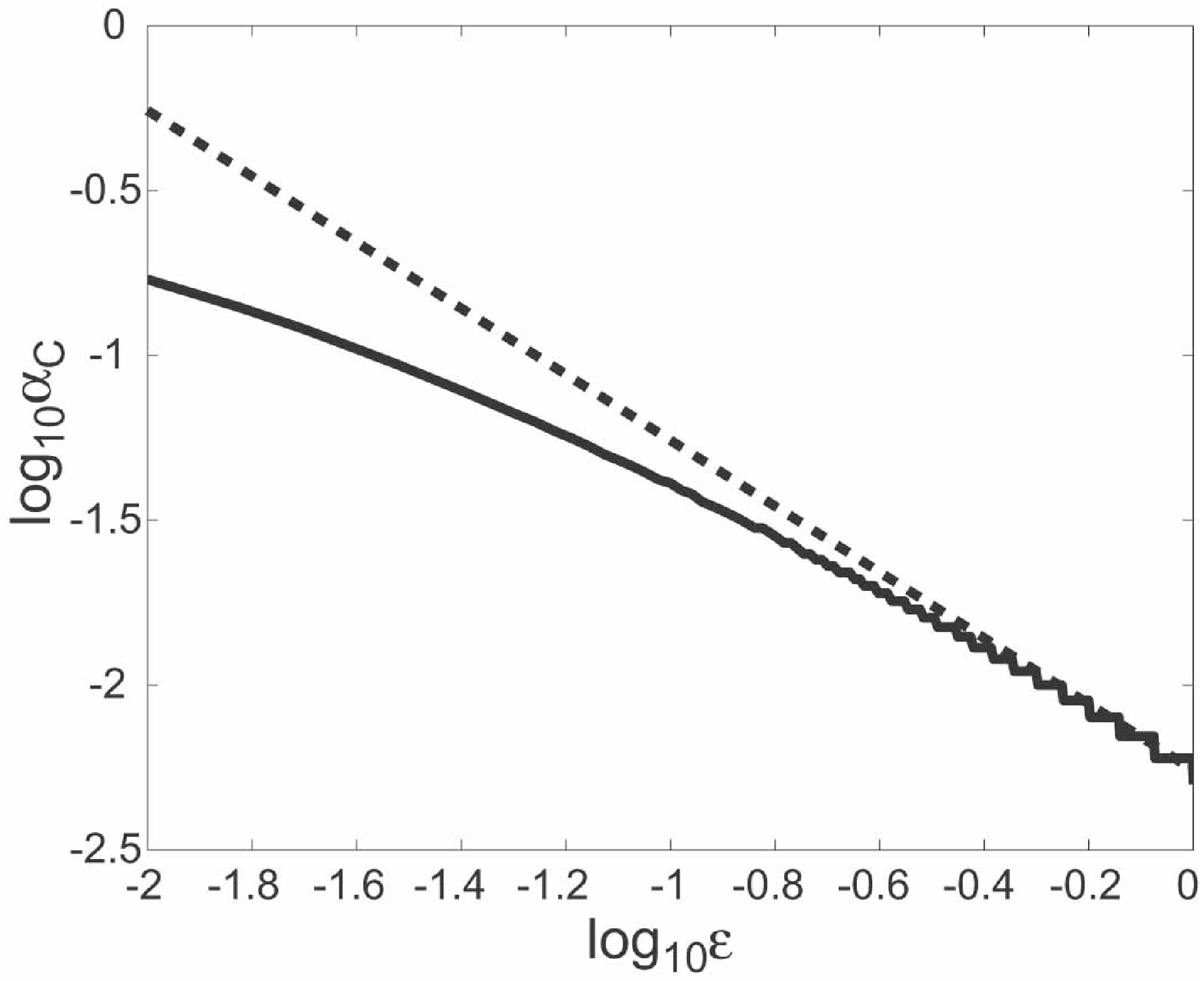}\\
\vspace{1.5cm}
{\Huge Figure 5}
\end{center}
\end{figure}

\end{document}